# A MODEL FOR REMOTE ACCESS AND PROTECTION OF SMARTPHONES USING SHORT MESSAGE SERVICE

K.S. Kuppusamy[1], Senthilraja.R [2], G. Aghila[3]

Department of Computer Science, School of Engineering and Technology,
Pondicherry University, Pondicherry, India

[1]`kskuppu@gmail.com`,[2]`rajasen006@gmail.com, aghilaa@yahoo.com`

*ABSTRACT*

*The smartphone usage among people is increasing rapidly. With the phenomenal growth of smartphone use, smartphone theft is also increasing. This paper proposes a model to secure smartphones from theft as well as provides options to access a smartphone through other smartphone or a normal mobile via Short Message Service. This model provides option to track and secure the mobile by locking it. It also provides facilities to receive the incoming call and sms information to the remotely connected device and enables the remote user to control the mobile through SMS. The proposed model is validated by the prototype implementation in Android platform. Various tests are conducted in the implementation and the results are discussed.*

## 1. INTRODUCTION

Nowadays, usage of mobile has become a vital part of day-to-day activities of people. We can refer the current time as the era of Smartphones. Suppressing all other traditional communication purpose, smartphones are now at the peak of popularity in their usage of accessing the internet which includes mail access, social networking, mobile shopping and mobile banking. Smartphone usage of people is studied in [1].

Smartphones contains critical and sensitive data of user like automated call records, photos, videos and saved passwords of WebPages. So loosing the smartphone means a very high amount of irrecoverable data loss which may not be affordable in many cases. Few surveys [2][3][4] about mobile theft in various countries has been studied. This claims the need of an intelligent application to be run in mobile to eradicate mobile theft and track the mobile even after change of the SIM also.

On the other hand, remote accessing of mobile becomes necessary when the mobile is left in somewhere like house or office. It includes getting the incoming call numbers, incoming messages, accessing call logs, changing phone's GPS, WIFI and profile settings and retrieving of contacts. The major objectives of the research work have been listed below.

- Locate the mobile and track it: The mobile location can be tracked using the proposed approach.

DOI : 10.5121/ijcseit.2012.2109   91

- Erase the critical data that has been stored in the mobile.
- Listen incoming calls and alert the remote device: The calls to the mobiles can be traced from a remote location.
- Listen incoming SMS and give automatic reply and/or forward to remote user: Similar to tracking calls the Short Messages can also be tracked.
- Access and change GPS, WIFI and profile settings through SMS.

The reminder of the paper is organised as follows: In section 2 the related works carried out in this area which has motivated this research work has been discussed. Section 3 deals about the proposed model and algorithm. Section 4 focuses on the implementation of the model in android platform and results of the implementation work. Sections 5 conclude the research paper and highlight the future implementation of this research work.

## 2. MOTIVATIONS

The mobile phones have evolved from simple communication devices to responsible units which can handle mission critical applications [5]. The security aspects of mobile devices are another important research area with many research models in it [6] [7] [8] [9].

The Smartphone related researches are plenty in number in the recent years. Norton Anti theft plug-in [10] enables the users to locate and lock the mobile through any web browser. It also provides facility to take picture through the camera of the smartphone and send to their online storage.

McAfee WaveSecure [11] also provides the similar functionality and it enables the users to erase their personal data in smartphone through online. These applications need computer to access smartphone and moreover these are concerned with mobile theft only.

The proposed model gives an innovative approach to access one smartphone through another smartphone or through a normal mobile. In addition to that, it provides the core functionalities of the applications mentioned above also.

This model uses SMS as the communication channel. Data that is flowing through SMS can be encrypted [12] and decrypted. This will secure the connection. The permission issues on Android also need to be considered while implementing the remote access [13].

## 3. PROPOSED MODEL

The proposed model has two components. One is server component that is dedicated to run in the smartphone which has to be accessed and protected. Another one is client component that should be in another smartphone to access the server component. In normal mobile phones, we can not install the client component. In this case, the users will communicate directly through SMS. So the server has been designed to handle the request from the client component as well as the normal SMS command from the mobile phones.

The following figure explains the model when the user is accessing his/her smartphone through other Smartphone. In this case, both should have the proposed application. The process starts from the client interface. To access a smartphone, the user should input the mobile number and the user defined remote connection command. This information will be encrypted by the client service and '$$' symbol will be appended in the first two character of the message and sent to the server. In server service, after reading the first two characters, the service will decrypt the data to verify the authentication information.

Once the authentication has been done, the server service will lock the mobile and display the login password prompt. So, the mobile becomes inaccessible thus secured. After this process, the request from the client service will be responded and encrypted by the server and it will sent back





to the client. The response will be decrypted in the client service and displayed in the client interface. All SMS from client service to server will be preceded by the '$$' symbol.

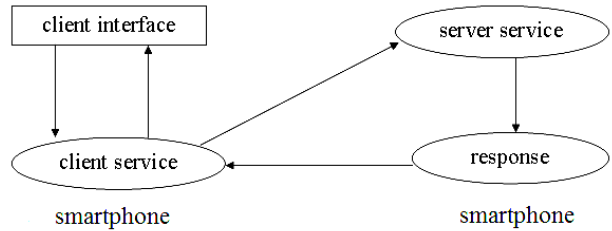

Figure 1: Overview of proposed model in a smartphone client.

The following figure demonstrates the model where the user wants to access the smartphone from a normal mobile. Here, there is no client component so the user has to message the '$' symbol followed by remote connection command from the mobile's messaging interface. Server service will distinguish the command from the smartphone's application and normal SMS by the first two character of the message. Depending on this, the response will be made. In this case, response won't be encrypted and it will be sent back as a SMS to the user. Response can be the contact information stored the smartphone, an incoming call number, etc,

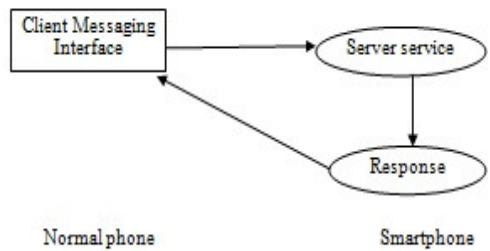

Figure 2: Overview of proposed model in a mobile client.

### 3.1 Architecture

The block diagram of the proposed model is shown in Figure 3. The SMS listener and parser in the server will receive all the incoming messages and check for the command in the SMS. In our proposed model, command will be preceded by the '$$' for the smartphone client and it will be preceded by the '$' for mobile client. If the SMS starts with the '$$' then the following characters (except first two) of the SMS will be decrypted. Authentication to remote mobile number has been done by checking the user defined login command in the received message. Once the authentication has done, the mobile keypad and touch screen will be locked and the prompt for the login password will be asked to unlock the mobile.

After this, SMS command received from the authenticated number will be sent to the server manager. There, the request handler will process the command and response will be made. Call handler will read the incoming call number and inform the client number by the call alert SMS if the call alert is activated. Boot handler will receive the mobile boot up so that the application will be started automatically when the mobile boots up. Mobile will be locked if the remote connection is active. Changing of sim card in that smartphone will be detected by the boot up listener and will be informed to remote user if it so. Database handler will do all the read and write operation of the database.

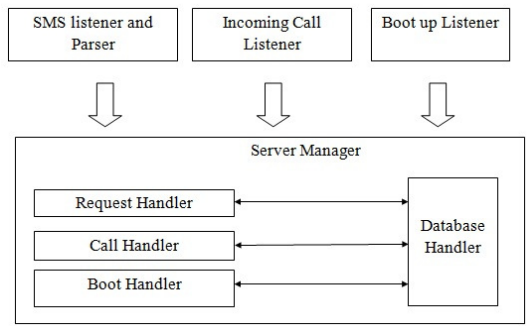

Figure 3: Architecture of server component.

Commands are the predefined words that will instruct the request handler to do a task. For example, $SILENT-ON is a command that will instruct the request handler to change the mobile sound to silent and $SILENT-OFF will change the phone profile to normal. All implemented commands are listed in the section 4.

The block diagram of the client component has been given in Figure 4. After sending the remote connection command, the SMS listener will monitor the inbox for the confirmation message from the server number. After getting confirmation message from the server number, the user will be allowed to access the further modules of the interface. This interface will provide options to send command to server number, search and retrieve contact, view the missed call list, and view inbox of the server number.

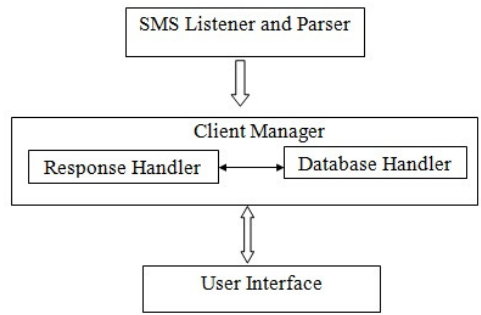

Figure 4: Architecture of client component.

### 3.3 Algorithm

The algorithm of the proposed model's client component has been given below.

Step 1: Get mobile number and its connection command from the user.
Step 2: Encrypt it and append '$$' in the starting and send to remote number.
Step 3: Wait for authentication confirmation.
Step 4: Create temporary login PIN (personal identification number) for the user to secure the client interface and show it to the user once after the authentication.
Step 5: Show the client interface to the user.

The algorithm of the proposed model's server component has been given below. It is having three sub modules. Algorithm for request listener has been given below.

Step 1: Receive incoming SMS.
Step 2: Check for the command SMS.
Step 3: Decrypt the message if it is encrypted.
Step 4: Check the database for the remote connection status.

Step 5: If a remote connection is not established then check for authentication. If it is an authenticated message then lock the mobile and send SMS to the user and go to step 9.
Step 6: Else check the incoming number. If message comes from authenticated number then go to step 8. Else go to next step.
Step 7: Perform SMS forwarding and automatic SMS reply if it is enabled by the remote user.
Step 8: Read the command from the message and perform it.
Step 9: Wait for next SMS.

The above algorithm is clearly depicted in the figure 5.

Algorithm for call handler of the server is given below.

Step 1: Monitor the incoming calls.
Step 2: If new call arrives, get the caller number.
Step 3: Check the call alert option in the database. If it is enabled go to next step.
Step 4: Send call alert SMS with mobile number and timestamp.

This algorithm is explained in Figure 6 (left side).

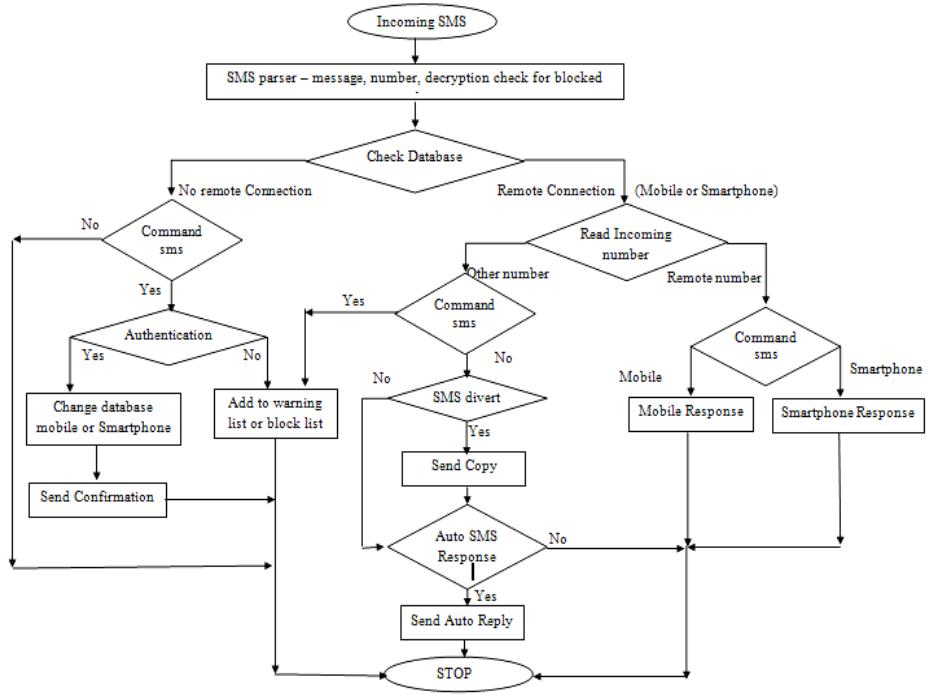

Figure 5: Server component – flow chart for request handler.

Algorithm for boot handler of the server is given below.

Step 1: Receive the boot up alert from the mobile framework.
Step 2: Check the database for the remote connection state.
Step 3: If remote connection enabled then lock the mobile and prompt for the login password.
Step 4: If phone's sim was changed then alert the remote user by sending message from the new number.

This algorithm is figured below (right side).

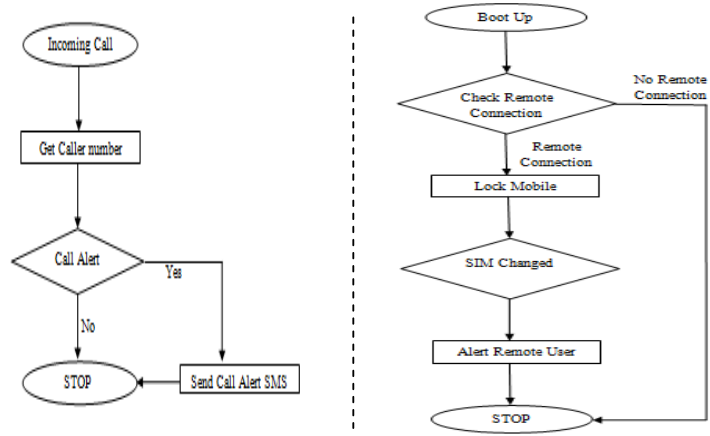

Figure 6: Server component – flow chart for call handler (left) & boot handler (right).

## 4. IMPLEMENTATION

The proposed model for smartphone security and remote access has been implemented in the Android 2.3 platform [14] Operating System. Client interface will help the user to send command by just clicking the radio buttons. This is show in the screen shots. But, normal mobile users have to type the command from the messaging interface of their mobile. The following commands are implemented.

Table 1: List of commands in the prototype implementation.

| COMMAND | DESCRIPTION |
| --- | --- |
| $SILENT-ON | Silent all the sound of the mobile. |
| $SILENT-OFF | Silent option will be removed. |
| $GPS-ON | Switch on the location finder and inform the user about the current location. It will work in background to track and inform the user often. |
| $GPS-OFF | Mobile tracking will be disabled. |
| $WIFI-ON | WIFI will be switched on by SMS command. |
| $WIFI-OFF | WIFI will be switched off to save battery usage. |
| $CALLALERT-ON | New incoming call details will be sent to remote user. |
| $CALLALERT-OFF | Call alert will be disabled. |
| $SMSDIVERT-ON | New incoming SMS's copy will be sent to remote user. |
| $SMSDIVERT-OFF | SMS divert will be disabled. |
| $SMS-REPLY ***** | Automatic SMS reply will be enabled. '*' are the message characters that will be sent as a reply. |
| $SMS-REPLY OFF | Automatic SMS reply will be disabled. |
| $CONTACT ***** | '*' are the character sequence of the contact name. If the contact has been found then its mobile and email address will be sent to the remote user. |
| $WIPEOUT | Clear all user data including memory card data. |
| $FLIGHT-ON | Flight mode will be activated to avoid incomings. |
| $SIGNOFF | Remote connection will be terminated. |

In the about set of commands, there was no command to initialize the remote connection. This is left to users so that they can create their own command. This will make the application more International Journal of Computer Science, Engineering and Information Technology (IJCSEIT), Vol.2, No.1, February 2012 secured and eliminates misuse of application by hacking it. For example, if the user is having command named "MYDOB" and activation pin "1989" then he has to type $MYDOB<space>1989. This will give the two way protection.

The Figure 7 contains three screen shots of the implemented model. The left side one provides option to the user to choose the client or the server component. The middle one will get the activation command, activation pin and login pin. This will be stored in database. First two information will be used to start the connection and last one i.e. login pin is used to terminate the remote connection directly in the smartphone. The right side one is the initial screen of the client component. The user has to give the mobile number and other information to start the connection.

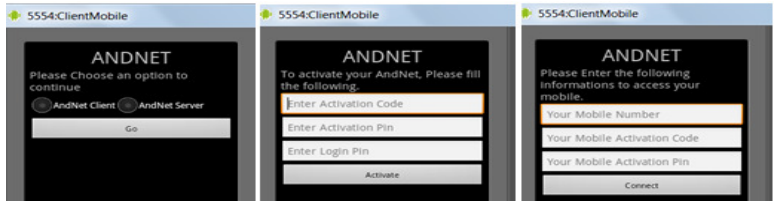

Figure 7: Screen shots of initial screens.

The Figures 8,9,10,11 are the screen shots of the implemented model in the android emulator [15] [16].

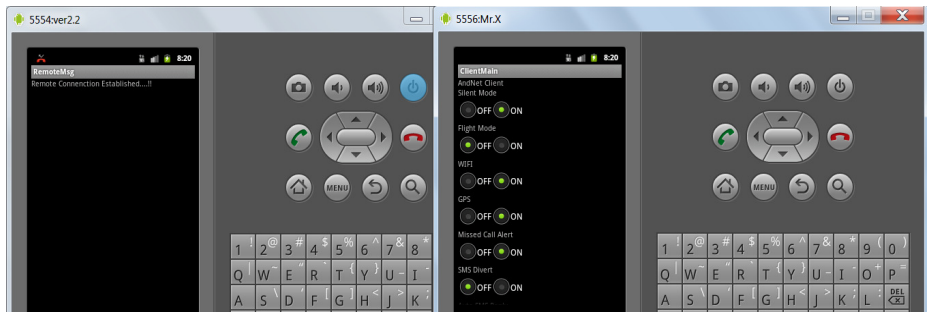

Figure 8: Client interface – sending command to smartphone.

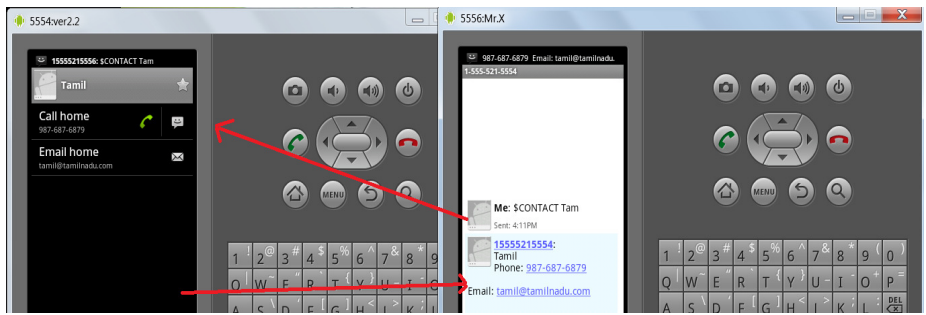

Figure 9: user retrieving contact detail through SMS.

97

International Journal of Computer Science, Engineering and Information Technology (IJCSEIT), Vol.2, No.1, February 2012

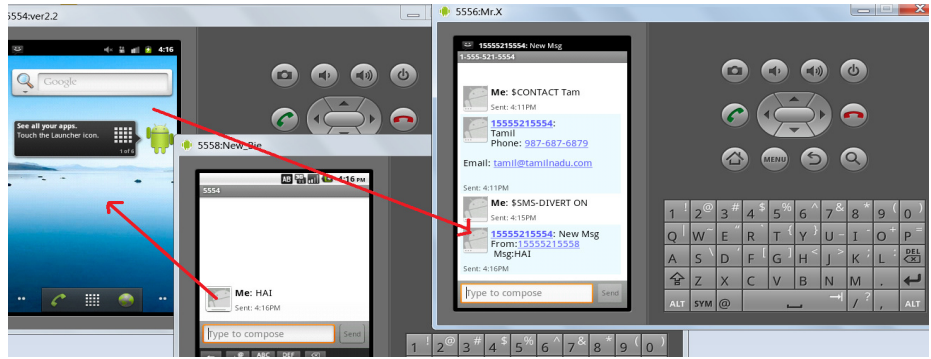

Figure 10: SMS divert – Mobile client getting diverted SMS.

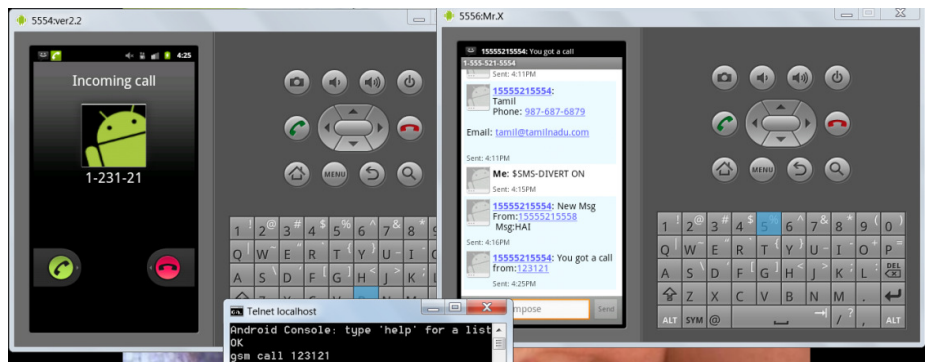

Figure 11: Call alert – Mobile client getting call alert.

Few tests were conducted in the implemented model. This helps to improve the model. To prevent misuse of the application, the model is updated with the following:

- Number that failed to provide the correct authentication data will be added in the warning list for the first two attempts. This data will exist in the list for 48hours.
- Within 48 hours, if third attempt also fails, then the number will be added in permanent block list. Options provided to user to delete the list.
- During activate remote connection, if command comes from other number then remote user will be alerted and requested number will be added to block list immediately.
- Default application installation path is set to mobile storage so that wiping out of memory card won't affect the application.
- SMS parser is improved. It will detect request from mobile number exactly. Request from websites like way2sms [17] will be rejected.
- Self request to remote connection (client number is same as the server) will be rejected.

## 5. CONCLUSION AND FUTURE DIRECTIONS

The proposed model has been implemented in android operating system. It was tested in Samsung galaxy pro smartphone. This provides the encouraging result. The model can be implemented in other smartphone platforms like windows, apple, etc. The conclusions drawn from the proposed system are listed below:

- The proposed model facilitates accessing of the device from a remote location using any other mobile terminal.

98

- The system has been designed in such a way that the mobile terminal used for accessing the remote android device, need not be an android device.

The future directions for this research work are listed below:

- The remote connection through SMS can be replaced by GPRS.
- The screen capturing of remote device can be incorporated so that the exact display can be accessed.

**REFERENCES**

[1] Hossein Falaki, Ratul Mahajan Srikanth Kandula, Dimitrios Lymberopoulos, Ramesh Govindan, Deborah Estrin: Diversity in smartphone usage, Proceedings of the 8th international conference on Mobile systems, applications, and services, ISBN: 978-1-60558-985-5

[2] Survey about mobile theft in UK: http://news.bbc.co.uk/2/hi/uk_news/1748258.stm

[3] Survey about mobile theft in USA: http://www.symantec.com/about/news/release/article.jsp?prId=20110208_01

[4] Survey about mobile theft in India:http://asiarelease.asia/norton-survey-reveals-1-in-2-indians-is-a-victim-ofmobile-phone-loss-or-theft/

[5] Patrick Traynor et.al, "From mobile phones to responsible devices" in "Security and Communication Networks", Wiley Publications, Vol 4 , Issue 6, 2011

[6] Karsten Sohr, Tanveer Mustafa, and Adrian Nowak. 2011. Software security aspects of Java-based mobile phones. In Proceedings of the 2011 ACM Symposium on Applied Computing (SAC '11). ACM, New York, NY, USA.

[7] W. Enck, M. Ongtang, and P. McDaniel. Understanding Android Security. IEEE Security and Privacy, 7(1):50–57, 2009.

[8] G. McGraw. Software Security: Building Security In. Addison-Wesley, 2006

[9] Bo Li and Eul Gyu Im: Smartphone, promising battlefield for hackers, Journal of Security Engineering , vol: 8 no: 1, 2011, pages 89-110

[10] Semantec Anti theft: https://market.android.com/details?id=com.symantec.anti.theft

[11] Mcafee Wave secure: https://market.android.com/details?id=com.wsandroid

[12] Kyungwhan Park, Gun Il Ma, Jeong Hyun Yi, Youngseob Cho, Sangrae Cho, Sungeun Park: Smartphone Remote Lock an d Wipe System with Integrity Checking of SMS Notification, Consumer Electronics (ICCE), IEEE International Conference on 9-12 Jan. 2011 pages 263-264.

[13] Adrienne Porter Felt, Erika Chin, Steve Hanna, Dawn Song, and David Wagner. 2011. Android permissions demystified. In Proceedings of the 18th ACM conference on Computer and communications security (CCS '11). ACM, New York, NY, USA, 627-638

[13] Android SDK: http://developer.android.com/sdk/android-2.3.html

[14] Android Emulator: http://developer.android.com/guide/developing/tools/emulator.html

[15] Operating System: http://en.wikipedia.org/wiki/Android_%28operating_system%29

[16] Online SMS sending portal: http://en.wikipedia.org/wiki/Way2SMS.com

**Authors**

**K.S.Kuppusamy** is an Assistant Professor at Department of Computer Science, School of Engineering and Technology, Pondicherry University, Pondicherry, India. He has obtained his Masters degree in Computer Science and Information Technology from Madurai Kamaraj University. He is currently pursuing his Ph.D in the field of Intelligent Information Management. His research interest includes Web Search Engines, Semantic Web.

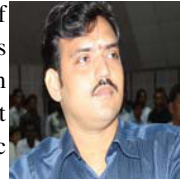

**Senthilraja . R** is a student pursuing Masters degree in computer applications at Department of Computer Science, School of Engineering and Technology, Pondicherry University, Pondicherry, India. He has completed his Bachelor degree in Computer Science from A.V.C College (Autonomous), Affiliated to Bharathidasan University in the year 2009 and he was selected as the best outgoing student of the college in the same year.

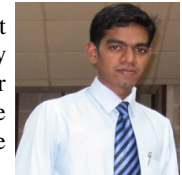

**G. Aghila** is a Professor at Department of Computer Science, School of Engineering and Technology, Pondicherry University, Pondicherry, India. She has got a total of 20 years of teaching experience. She has received her M.E (Computer Science and Engineering) and Ph.D. from Anna University, Chennai, India. She has published more than 60 research papers in web crawlers, ontology based information retrieval. She is currently a supervisor guiding 8 Ph.D. scholars. She was in receipt of Schrneiger award. She is an expert in ontology development. Her area of interest include Intelligent Information Management, artificial intelligence, text mining and semantic web technologies

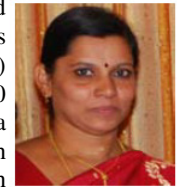